\documentclass[showpacs,preprintnumbers,
superscriptaddress, groupedaddress,prb]{revtex4}
\usepackage{amssymb}
\usepackage{amsmath}
\usepackage{graphicx}
\usepackage{dcolumn}
\usepackage{bm}

\begin{document}

\title{ Many-Body Correlation Effects in the Spatially Separated Electron
and Hole Layers in the Coupled Quantum Wells}
\author{V.S. Babichenko}
\affiliation{RNC Kurchatov Institute, Kurchatov Sq.1, 123182, Moscow, Russia}
\author{I. Ya. Polishchuk}
\affiliation{Max Planck Institute for the Physics of Complex Systems, N\"{o}thnitzer Str.
38, D-01187 Dresden, Germany}
\affiliation{RNC Kurchatov Institute, Kurchatov Sq.1, 123182, Moscow, Russia}
\date{\today }

\begin{abstract}
The many-body correlation effects in the spatially separated electron and
hole layers in the coupled quantum wells (CQW) are investigated. The
specific case of the \textit{many-component }electron-hole system is
considered, $\nu \gg 1$ being the number of components. Keeping the main
diagrams in the parameter $1/\nu $ allows us to justify the $RPA$ diagram
selection. The ground state of the system is found to be the \textit{%
electron-hole liquid} which possesses the energy smaller than the exciton
phase. The connection is discussed between the results obtained and the
experiments in which the inhomogeneous state in the CQW is revealed.
\end{abstract}

\maketitle

\section{Introduction}

The investigation of spatially separated electrons and holes in the coupled
quantum wells (CQW) is motivated by expecting that the electron-hole pairs
forming a long-living bound state, namely exciton, can experience the
Bose-Einstein condensation \cite{Lozovik}. The interest in such systems has
greatly grown in the recent years due to the increasing ability to
manufacture the high quality quantum well structures in which electrons and
holes are confined in the different regions between which the tunneling can
be made negligible \cite{Review2011}. For this reason, in the CQW the
exciton lifetime becomes by several orders of the magnitude longer than the
lifetime of excitons in bulk materials, enabling an experimental observation
of the exciton Bose-Einstein condensation.

As early as decade and a half, Lozovik and Berman, using the variational
approach, predicted the existence of an electron-hole condensate phase in
CQW \cite{Lozovik-Berman1996}. The further theoretical investigations of
such systems reveal that the phase diagram of the system can be rather rich
\cite{Balatsky}.

The main goal of the paper is to find out how the many-body correlations in
the electron-hole system in the CQW influence its ground state. A specific
case of the \textit{many-component }spatially separated electron-hole plasma
in the CQW is studied in the current paper. It is assumed that $\nu $
different kinds of electrons are confined in one layer of the CQW, while $%
\nu $ different kinds of holes are confined in the other layer, the
parameter being $\nu \gg 1.$ For the first time, the large $\nu $
approximation was proposed for the investigation of a 3D electron-hole
liquid in many-valley semiconductors \cite{babich}. Then, this approach got
a further development \cite{keldysh}.

In the model under consideration, the thickness of the layers is supposed to
be so small that the carriers are the $2D$ ones. Let $a_{B}$ be the
effective Bohr radius and $l$ be the inter-layer distance. Below, we mainly
consider the case $l\gg a_{B}.$ Under such condition, the isolated
electron-hole pair may form an exciton possessing the \textit{in-plane}
radius $R_{ex}\sim a_{B}\left( l/a_{B}\right) ^{3/4}\gg a_{B}$ \cite{Lozovik}%
. Let $n$ be the carrier concentration. The concept of excitons has sense
until the average in-plane inter-exciton distance being of the order of $%
n^{-1/2}$ is larger than $R_{ex},$ i.e. $R_{ex}^{2}n\ll 1.$ At higher
density the exciton system brakes down into the electron-hole one.

It is shown in this paper that the negative contribution of the energy
induced by the intra-layer carrier correlation strongly reduces the total
energy. This effect facilitates \textit{the electron-hole liquid} formation.
For this phase, the equilibrium carrier density $n_{eq},$which is the same
for the both layers, is found to obey the condition $R_{ex}^{2}n_{eq}\sim 1.$
Thus, the criterion for an existence of the exciton state is violated.
Therewith, the \textit{electron-hole liquid} is shown to possess the energy
\textit{lower} as compared to \textit{the} \textit{exciton state}. Note,
that the similar conclusion was made for conventional 3D semiconductors, as
well \cite{keldysh1, rice}.

The \textit{homogeneous }electron-hole liquid state predicted in this paper
is realized within the assumption that the carriers possess the infinite
lifetime. If not, the system breaks down into the \textit{low-density
exciton gas }phase and the \textit{dense electron-hole liquid} phase.
Between these phases the dynamic equilibrium is established which is
supported by the external pumping. A similar scenario was proposed by
Keldysh for conventional 3D semiconductors \cite{keldysh1}.

The diagrammatic technique is used to calculate the correlation contribution
to the energy of the electron-hole state. The approach is based on the $%
1/\nu $ expansion, what allows one to justify the random phase approximation
(RPA) diagram selection. It is interesting to note that, for $l\gg a_{B},$
the ground state energy as well as the equilibrium carrier concentration $%
n_{eq}$ do not depend on the parameter $\nu .$ This unexpected feature
allows us to expect that the results obtained may be, at least
qualitatively, extended even to the region $\nu \sim 1$.

The paper is organized as follows. First, we describe the model of the
electron-hole system in the CQW. Then, the self-consistent diagrammatic
approach is proposed to calculate the electron and hole Matsubara Green
functions and estimate the chemical potential $\mu $ for the electrons and
holes at temperature $T\ll \varepsilon _{F},$ $\varepsilon _{F}$ being the
Fermi energy for both electrons and holes. This allows us to find the
equilibrium \textit{electron-hole liquid state} concentration $n_{eq}$.
Though, to obtain the main results, it is assumed that $l\gg $ $a_{B},$the
case $l\leq a_{B}$ is also analyzed at the end of the paper. In conclusion,
we consider a possible relation between the results obtained and the
experiments, in which the nonuniform state of the system in CQW is revealed
\cite{Nature-2002-Butov, Nature-2002-Snoke, Butov2003, Snoke2003, Butov2004,
Larionov2002, Dremin2004, Timofeev2005}.

\section{Model and The Calculation of the Self-Energy}

To describe the spatially separated electron-hole system in the CQW, it is
supposed that the electrons are confined within one infinitely narrow $2D$
layer, while the holes are confined within the other layer. For the sake of
simplicity, it is supposed that the effective mass of both electrons and
holes are the same. Below the system of units is used with the effective
electron (hole) charge $e=1,$ effective electron (hole) masses $m=1,$ and
the Planck constant $\hbar =1$. Then, the effective Bohr radius is $a_{B}=1,$
and the energy is measured in the Hartree units.

Let $n$ be concentration of electrons or holes. According to the model under
consideration, the Fermi momentum $p_{F}=2\pi ^{1/2}\left( n/\nu \right)
^{1/2}$ and the Fermi energy $\varepsilon _{F}=2\pi n/\nu $ are the same for
both electrons and holes. We confine ourselves to the case of strongly
degenerated plasma, that is, the temperature $T\ll \varepsilon _{F}.$

The Hamiltonian of the system is $\widehat{H}=\widehat{H_{0}}+\widehat{U,}$ $%
H_{0}$ being the kinetic energy and $\widehat{U}$ being the Coulomb
interaction. In the second quantization one has 
\begin{equation}
\widehat{H_{0}}=\sum_{\alpha \sigma \mathbf{k}}\frac{k^{2}}{2}a_{\alpha
\sigma }^{+}\left( \mathbf{k}\right) a_{\alpha \sigma }\left( \mathbf{k}%
\right) ;\widehat{U}=\frac{1}{2V}\sum_{\alpha \alpha ^{\prime }\sigma \sigma
^{\prime }\mathbf{kk}^{\prime }\mathbf{q}}U_{\alpha \alpha ^{\prime }}\left(
\left\vert \mathbf{k}\right\vert \right) a_{\alpha \sigma }^{+}\left(
\mathbf{k}\right) a_{\alpha ^{\prime }\sigma ^{\prime }}^{+}\left( \mathbf{k}%
^{\prime }\right) a_{\alpha ^{\prime }\sigma ^{\prime }}\left( \mathbf{k}%
^{\prime }-\mathbf{q}\right) a_{\alpha \sigma }\left( \mathbf{k}+\mathbf{q}%
\right) .  \label{eq1.1}
\end{equation}

%
Here $\alpha =e$ for the electrons and $\alpha =h$ for the holes. Then, $%
a_{\alpha \sigma }^{+}\left( \mathbf{k}\right) $ and $a_{\alpha \sigma
}\left( \mathbf{k}\right) $ are the creation and annihilation operators, $%
\mathbf{k}$ is the in-plane $2D$ momentum, $\sigma =1,...\nu $ denotes the
kind of the electron or the hole component. The Coulomb interaction in the
momentum representation reads
\begin{equation}
U_{\alpha \alpha ^{\prime }}\left( \left\vert \mathbf{k}\right\vert \right)
=\left\{
\begin{array}{c}
\frac{2\pi }{k},\alpha =\alpha ^{\prime } \\
-\frac{2\pi }{k}e^{-kl},\alpha \neq \alpha ^{\prime }%
\end{array}%
\right. .  \label{eq2}
\end{equation}

\begin{figure}[tbph]
\centering
\includegraphics[width=0.4\textwidth]{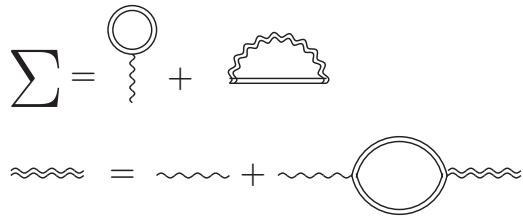} 
\caption{The self-consistent diagrammatic equations for the Green function
and mass operator. Here ${\Sigma }$ is the mass operator. The double solid
line denotes the total Green function $G.~$ The wavy line represents the
interaction (\protect\ref{eq2}) and the double wavy line corresponds to the
renormalized interaction.}
\label{Figure2}
\end{figure}

Let $G_{\alpha ,\sigma }^{0}\left( \omega ,\mathbf{p}\right) =\left( i\omega
+\mu -\varepsilon _{p}\right) ^{-1}$ be the electron (hole) free Matsubara
Green function and $G_{\alpha ,\sigma }\left( \omega ,\mathbf{p}\right)
=\left( i\omega +\mu -\Sigma _{\alpha ,\sigma }\left( \omega ,\mathbf{p}%
\right) -\varepsilon _{p}\right) ^{-1}~$be the electron (hole) total Green
function. Here $\omega $ is the Matsubara frequency, $\varepsilon
_{p}=p^{2}/2,$ $\Sigma _{\alpha ,\sigma }\left( \omega ,\mathbf{p}\right) $
is the mass operator, and $\mu $ is the chemical potential. Let us consider
the system of self-consistent equations whose diagram representation is
shown in Fig.\ref{Figure2}

The first diagrammatic equation in Fig.\ref{Figure2} reads
\begin{eqnarray}
\Sigma _{\alpha ,\sigma }\left( \varepsilon ,p\right) &=&\Sigma _{\alpha
,\sigma }^{(1)}\left( \varepsilon ,p\right) +\Sigma _{\alpha ,\sigma
}^{(2)}\left( \varepsilon ,p\right) ,  \label{eq2.01} \\
\Sigma _{\alpha ,\sigma }^{(1)}\left( \varepsilon ,p\right) &=&\frac{T}{V}%
\sum_{\alpha ^{\prime }\sigma ^{\prime }\mathbf{k}\omega }U_{\alpha \alpha
^{\prime }}\left( 0\right) G_{\alpha ^{\prime }\sigma ^{\prime }}\left(
\mathbf{k},\omega \right) ,  \label{eq2.02} \\
\Sigma _{\alpha ,\sigma }^{\left( 2\right) }\left( \varepsilon ,p\right) &=&-%
\frac{T}{V}\sum_{\mathbf{k}\omega }\widetilde{U}_{\alpha \alpha }\left(
\mathbf{k},\omega \right) G_{\alpha \sigma }\left( \mathbf{p}+\mathbf{k}%
,\varepsilon +\omega \right) .  \label{eq2.03}
\end{eqnarray}%
Here $T$ is the temperature, and $V$ is the square of the layers.

The renormalized interaction $\widetilde{U}_{\alpha \alpha }\left( \mathbf{p}%
,\omega \right) $ in Eq. (\ref{eq2.03}) obeys the second diagrammatic
equation in Fig.\ref{Figure2} and reads
\begin{equation}
\widetilde{U}_{\alpha \alpha }\left( \mathbf{p},\omega \right) =U_{\alpha
\alpha }\left( \mathbf{p}\right) +U_{\alpha \alpha }\left( \mathbf{p}\right)
\Pi _{\alpha \sigma }\left( \mathbf{p},\varepsilon \right) \widetilde{U}%
_{\alpha \alpha }\left( \mathbf{p},\omega \right) .  \label{eq2.04}
\end{equation}%
The polarization operator%
\begin{equation}
\Pi \left( \mathbf{p},\varepsilon \right) =\Pi _{\alpha \sigma }\left(
\mathbf{p},\varepsilon \right) =\frac{T}{V}\sum\limits_{\mathbf{k}\omega
}G_{\alpha \sigma }\left( \mathbf{p}+\mathbf{k,}\varepsilon +\omega \right)
G_{\alpha \sigma }\left( \mathbf{k},\omega \right)  \label{eq2.3}
\end{equation}%
does not depend on the kind of the particles.

Let us explain the rule of diagram selection. Each fermion loop contributes
a factor $\nu $ to the diagrams. Among all the diagrams of the given order,
only the diagrams are retained which contain the \textit{maximal} number of
the fermion loops. Thus, our approach is a $1/\nu $ expansion which results
in keeping only the diagram of the \textit{RPA kind}. The diagrams for which
the bubbles are linked by more than one interaction line are small as
compared with the RPA ones in the parameter $1/\nu .$For the same reason,
the diagrams are small, which contain any \textit{vertex corrections}.

The expression for the renormalized interaction follows from Eq. (\ref%
{eq2.04})
\begin{equation}
\widetilde{U}_{\alpha \alpha }\left( \mathbf{p},\omega \right) =\widetilde{U}%
\left( \mathbf{p},\omega \right) =\frac{U_{\alpha \alpha }\left( \mathbf{p}%
\right) }{1-\nu \cdot U_{\alpha \alpha }\left( \mathbf{p}\right) \Pi \left(
\mathbf{p},\omega \right) }.  \label{eq2.2}
\end{equation}%
Let us note that, like the polarization operator $\Pi _{\alpha \sigma
}\left( \mathbf{p},\omega \right) $, both the mass operator $\Sigma _{\alpha
,\sigma }\left( \varepsilon ,p\right) $ and the renormalized interaction $%
\widetilde{U}_{\alpha \alpha }\left( \mathbf{p},\omega \right) $ do not
depend on the kind of the particle.

Our first goal is to calculate the chemical potential $\mu $ as a function
of the density $n$ using the self-consistent system of Eqs. (\ref{eq2.01})-(%
\ref{eq2.2}). Let us remind the exact relation

\begin{equation}
\mu -\Sigma \left( 0,p_{F}\right) =p_{F}^{2}/2.  \label{eq2.21}
\end{equation}%
Then, the convergence method is used. If the interaction is neglected, the
mass operator $\Sigma \left( 0,p_{F}\right) \approx 0$ and, therefore,
\begin{equation}
\mu \approx \mu _{0}=p_{F}^{2}/2=2\pi n/\nu .  \label{eq2.211}
\end{equation}

To find the next iteration for the chemical potential, let us calculate $%
\Sigma \left( 0,p_{F}\right) $ taking $\mu =\mu _{0}$. In this case, one
should replace $G$ by $G_{0}$ in the above system of equations. First,
taking into account Eq. (\ref{eq2}), we obtain
\begin{equation}
\Sigma _{\alpha \sigma }^{(1)}\left( \varepsilon ,\mathbf{p}\right) =\frac{T%
}{V}\sum_{\alpha ^{\prime }\sigma ^{\prime }\mathbf{k}\omega }U_{\alpha
\alpha ^{\prime }}\left( 0\right) G_{\alpha ^{\prime }\sigma ^{\prime
}}^{0}\left( \mathbf{k},\omega \right) =2\pi ln.  \label{eq2.2111}
\end{equation}

Then, to find the next contribution to the mass operator $\Sigma _{\alpha
\sigma }^{2}\left( \omega ,\mathbf{p}\right) ,$ one should first estimate
the renormalized interaction $\widetilde{U}_{\alpha \alpha }\left( \mathbf{k}%
,\omega \right) $. For this purpose, let us consider the polarization
operator in zero order in the interaction:

\begin{equation}
\Pi \left( \mathbf{p},\varepsilon \right) \approx \frac{1}{\left( 2\pi
\right) ^{2}}\int d^{2}k\frac{n_{0}\left( \mathbf{k}\right) -n_{0}\left(
\mathbf{k+p}\right) }{i\hbar \varepsilon +\mathbf{k}^{2}/2-\left( \mathbf{k+p%
}\right) ^{2}/2},  \label{eq3}
\end{equation}%
where $n_{0}\left( \mathbf{k}\right) =\left[ \exp \frac{\left( k^{2}/2-\mu
_{0}\right) }{T}+1\right] ^{-1}$ is the Fermi distribution function.

For \textit{small }momenta and frequencies $p\lesssim p_{F},$ $\varepsilon
\lesssim \varepsilon _{F},$ one finds $\Pi _{0}\left( \mathbf{p},\varepsilon
\right) \simeq -1/2\pi .$ Substituting this value into Eq. (\ref{eq2.2})
gives the estimate
\begin{equation}
\widetilde{U}\left( \mathbf{p},\varepsilon \right) =\frac{U_{\alpha \alpha
}\left( \mathbf{p}\right) }{1+U_{\alpha \alpha }\left( \mathbf{p}\right) \nu
\frac{1}{2\pi }}\simeq \frac{1}{\nu }\ll 1.  \label{eq3.11}
\end{equation}%
This estimation is valid if

\begin{equation}
n^{1/3}/\nu \ll 1,  \label{eq01}
\end{equation}%
what is supposed below.

Let us turn to the behavior of the renormalized interaction $\widetilde{U}%
\left( \mathbf{p},\varepsilon \right) $ for \textit{\ large} momenta \textbf{%
\ }$\left\vert \mathbf{p}\right\vert \gg p_{F}.$ First, let us estimate the
polarization operator (\ref{eq3}) for such momenta. Consider, for example,
the contribution $\Pi ^{\prime }\left( \mathbf{p},\varepsilon \right) $
associated with the factor $n_{\alpha \sigma }\left( \mathbf{p}+\mathbf{q}%
\right) .$ It is evident that momenta $\mathbf{k}$ alone contribute to the
polarization operator for which $\left\vert \mathbf{p}+\mathbf{k}\right\vert
\leq p_{F}.$ For this reason, \textbf{\ }$\left\vert \mathbf{k}\right\vert
\gg p_{F}.$ Therefore, $\left( \mathbf{p}+\mathbf{k}\right) ^{2}\ll \mathbf{k%
}^{2}\approx \mathbf{p}^{2}$ and
\begin{equation*}
\Pi ^{\prime }\left( \mathbf{p},\varepsilon \right) \approx -\frac{1}{\left(
2\pi \right) ^{2}}\frac{n/\nu }{i\hbar \varepsilon +\mathbf{p}^{2}/2}.
\end{equation*}%
Similarly, one can estimate the contribution $\Pi _{0}^{\prime \prime
}\left( \mathbf{p},\varepsilon \right) $ connected with the term $n_{\mathbf{%
p}}$ in Eq. (\ref{eq3}). As a result, we obtain
\begin{equation}
\Pi \left( \mathbf{p},\varepsilon \right) \approx -\frac{1}{\left( 2\pi
\right) ^{2}\nu }\frac{n\mathbf{p}^{2}}{\varepsilon ^{2}+\left( \mathbf{p}%
^{2}/2\right) ^{2}},~p\gg p_{F}.  \label{eq4}
\end{equation}%
Taking into account (\ref{eq4}), one can estimate the renormalized
interaction (\ref{eq2.2})

\begin{equation}
\widetilde{U}\left( \mathbf{p},\omega \right) =\frac{\frac{2\pi }{p}}{1+%
\frac{2\pi }{p}\frac{np^{2}}{\omega ^{2}+\left( p^{2}/2\right) ^{2}}},~p\gg
p_{F}.  \label{eq4.2}
\end{equation}%
Substituting (\ref{eq3.11}) into (\ref{eq2.03}) gives the estimate for the
contribution to the self-energy $\Sigma _{\alpha \sigma }^{\left( 2\right)
}\left( \omega ,\mathbf{p}\right) $ coming from $\left\vert \mathbf{k}%
\right\vert \leq p_{F}.$
\begin{equation}
\left\vert \Sigma _{\alpha ,\sigma }^{\prime \left( 2\right) }\left(
\varepsilon ,p\right) \right\vert \approx \frac{1}{\pi }\frac{n}{\nu ^{2}}.
\label{eq4.201}
\end{equation}%
Let us turn to the contribution to $\Sigma _{\alpha \sigma }^{\left(
2\right) }\left( \omega ,\mathbf{p}\right) $ gained from $\left\vert \mathbf{%
k}\right\vert \gg p_{F}.$ It is convenient to represent it in the form%
\begin{gather}
\Sigma _{\alpha ,\sigma }^{\prime \prime \left( 2\right) }\left( \varepsilon
,p\right) =-\frac{T}{V}\sum_{\left\vert \mathbf{k}\right\vert \gg
p_{F},~\omega }(U\left( \mathbf{k}\right) G_{\alpha \sigma }^{0}\left(
\mathbf{p}+\mathbf{k},\varepsilon +\omega \right) -  \notag \\
-\frac{T}{V}\Delta U\left( \mathbf{k}\right) G_{\alpha \sigma }^{0}\left(
\mathbf{p}+\mathbf{k},\varepsilon +\omega \right) ).  \label{eq4.21}
\end{gather}%
Here $\Delta U\left( \mathbf{k}\right) =\widetilde{U}\left( \mathbf{k}%
,\omega \right) -U\left( \mathbf{k}\right) .$ The first sum on the R.H.S. in
Eq. (\ref{eq4.21}) can be estimated as

\begin{equation}
\frac{1}{\left( 2\pi \right) ^{2}}\int_{\left\vert \mathbf{k}\right\vert \gg
p_{F}}\frac{d^{2}\mathbf{k}}{\left\vert \mathbf{k}\right\vert }n_{0}\left(
\mathbf{k+p}\right) \leq \frac{1}{\pi ^{3/2}}\left( \frac{n}{\nu }\right)
^{1/2}  \label{eq4.22}
\end{equation}

Let us turn to the second term on the R.H.S. in Eq. (\ref{eq4.21}). Since we
are interested in $\Sigma _{\alpha ,\sigma }^{\prime \prime 2}\left(
0,p_{F}\right) ,$ let us substitute $\left[ G_{\alpha \sigma }^{0}\left(
\mathbf{p}+\mathbf{k},\varepsilon +\omega \right) +G_{\alpha \sigma
}^{0}\left( \mathbf{p}-\mathbf{k},\varepsilon -\omega \right) \right]
\approx $ $-\frac{k^{2}}{\omega ^{2}+\left( k^{2}/2\right) ^{2}}\left( k\gg
p_{F}\right) $ into this term. Then, one obtains the estimate for the second
sum in Eq. (\ref{eq4.21})

\begin{equation}
-\frac{1}{\left( 2\pi \right) ^{3}}\int d\omega \int_{k\gg p_{F}}kdk\frac{%
\left( \frac{2\pi }{k}\right) ^{2}n\left( \frac{k^{2}}{\omega ^{2}+\left(
k^{2}/2\right) ^{2}}\right) ^{2}}{1+\frac{2\pi }{k}\frac{nk^{2}}{\omega
^{2}+\left( k^{2}/2\right) ^{2}}}\approx -Cn^{1/3}.  \label{eq4.31}
\end{equation}%
This estimate results from the following hint. Let us introduce new
variables $\omega /n^{2/3}$ and $k=k/n^{1/3}$. The integrand is maximal in
the vicinity of the region $\omega /n^{2/3}$ $\sim k/n^{1/3}\sim 1.$ Due to
Eq. (\ref{eq01}), one can put the lower limit equal to zero in integral (\ref%
{eq4.31}). Then the integral depends on the concentration as $n^{1/3}.$ The
proportionality constant is calculated numerically, $C\approx 0.615$.

\section{Equation of state}

Taking into account condition (\ref{eq01}), one can neglect the contribution
of expressions (\ref{eq2.211}) (\ref{eq4.201}), (\ref{eq4.22}) into the
chemical potential as compared with (\ref{eq4.31}). Thus, expressions (\ref%
{eq2.2111}) and (\ref{eq4.31}) alone contribute to the chemical potential.
Therefore,
\begin{equation}
\mu \approx 2\pi ln-Cn^{1/3}.  \label{eq5}
\end{equation}%
The chemical potential $\mu ,$ given by Eq. (\ref{eq5}), is a result of the
first iteration. Let us now substitute expression (\ref{eq5}) for the
chemical potential into Eqs. (\ref{eq2.01}) - (\ref{eq2.03}) instead of $\mu
_{0}$ and repeat all the calculations starting from Eq. (\ref{eq2.2111}). It
is easy to find that, whether $\mu _{0}$ or $\mu $ is taken as a starting
point, the correlation contribution to the self-energy is of the same order
of the magnitude given by (\ref{eq4.31}). For this reason, the expression
for the chemical potential (\ref{eq5}) is the stable solution of the
self-consistent system of equations, the constant $C$ being of the order of
unity. This chemical potential corresponds to the energy per unit volume%
\begin{equation}
\varepsilon =\pi ln^{2}-\frac{3}{4}Cn^{4/3}.  \label{eq50}
\end{equation}%
Then, at \textit{zero temperature }the pressure is
\begin{equation}
P=\mu n-\varepsilon =\pi ln^{2}-\frac{C}{4}n^{4/3}.  \label{eq51}
\end{equation}%
At small density $n<$ $\left( C/6\pi l\right) ^{3/2},$ the compressibility $%
\partial P/\partial V>0,$ and the system is unstable. For higher density, in
the interval $\left( C/6\pi l\right) ^{3/2}<n<\left( C/4\pi l\right) ^{3/2}$%
, one has $\partial P/\partial V<0.$ The system is stable and has a \textit{%
negative} pressure. The liquid state is reached at the upper limit of this
interval
\begin{equation}
n_{eq}=\left( C/4\pi l\right) ^{3/2}.  \label{eq6}
\end{equation}%
where the pressure $P=0.$

Thus, for the electron-hole liquid in the equilibrium, the bound energy%
\textit{\ per one particle }is
\begin{equation}
\varepsilon _{eq}=-\varepsilon \left( n_{eq}\right) /n_{eq}=-\frac{C^{3/2}}{%
4\pi ^{1/2}}\frac{1}{l^{1/2}}.  \label{eq7}
\end{equation}%
At the same time, for the \textit{isolated exciton }the bound energy $%
\varepsilon _{ex}^{0}~$is known to behave as $\varepsilon _{ex}^{0}\sim
-1/l. $ Since the excitons in the CQW experience\textit{\ repulsion,.} the
bound energy per particle is $\varepsilon _{ex}>-1/l.$ Comparing this energy
with $\varepsilon _{eq}$ (see expression (\ref{eq7})) one concludes that the
electron-hole state possesses a smaller energy as compared to the exciton
state for $l\gg 1$. In addition, note that the exciton in-plane radius $%
R_{ex}\sim l^{3/4}$ and, thus, for the equilibrium carrier concentration
obtained, one has $n_{eq}R_{ex}^{2}\sim 1.$ Under such conditions, the
exciton concept seems us to have no sense.

\section{Discussion and Conclusion}

The results above are obtained for the case $l\gg 1,$ i.e., $l\gg a_{B}.$ In
this case, the ground-state of the system is the electron-hole liquid which
possesses the equilibrium density (\ref{eq6}). This is a stable state which
has the energy smaller than that of the exciton state.

Consider, however, the case $l\ll 1.$ It follows from Eqs. (\ref{eq2.211}), (%
\ref{eq2.2111}), (\ref{eq4.31}) that the kinetic energy $\varepsilon _{kin}$%
, the direct Coulomb interaction $\varepsilon _{eh}$ and the intra-layer
correlation energy $\varepsilon _{corr}$ can be estimated as
\begin{equation*}
\varepsilon _{kin}\sim \frac{n}{\nu },\varepsilon _{e-h}\sim nl,\varepsilon
_{corr}\sim -n^{1/3},
\end{equation*}%
respectively. The equilibrium is reached at the concentration
\begin{equation}
n_{eq}\sim \left( \frac{\nu }{1+\nu l}\right) ^{3/2},  \label{eq8-0}
\end{equation}%
for which the total energy $\varepsilon _{eq}$ is minimal and
\begin{equation}
\varepsilon _{eq}\sim -\left( \frac{\nu }{1+\nu l}\right) ^{1/2}\approx
\left\{
\begin{array}{c}
-\frac{1}{l},\text{if~}\nu l\gg 1 \\
-\nu ^{1/2},\text{if~}\nu l\ll 1%
\end{array}%
\right. .  \label{eq8}
\end{equation}%
For the case $l\ll 1$ considered in this place, the exciton bound energy $%
\varepsilon _{ex}\sim -1$ in the units taken in the paper. (In fact, this
energy is even larger due to repulsion between the excitons). Throughout the
paper it is assumed that $\nu \gg 1.$Thus, $\varepsilon _{eq}\ll \varepsilon
_{ex},$ as well as above in the case $l\gg 1$. Also, for the concentration (%
\ref{eq8-0}), the average distance between the carriers is of the order of $%
~n_{eq}^{-1/2}\sim $ $\left( \frac{1+\nu l}{\nu }\right) ^{3/4}\ll 1.$ At
the same time, for $l\ll 1$ the exciton in-plane radius $R_{ex}\sim a_{B}=1.$
Thus, the exciton in-plane radius $R_{ex}$ exceeds the average distance
between the carriers, what has no sense. Thus, both in the case $l\gg 1$ and
in the case $l\ll 1$ the ground state of the system of spatially separated
electrons and holes in the CQW is the electron-hole liquid.

As the intermediate case $l\sim 1$ is concerned, both $\varepsilon _{eq}$ $%
\sim -1$ and $\varepsilon _{ex}\sim -1.$ However, in this case $%
n_{eq}^{-1/2}\sim R_{ex}\sim a_{B}\sim 1$ and, thus, the exciton concept
seems has no sense as well.

Let us return to the case $l\gg 1.$An interesting feature of the results
obtained is that both the bound energy (\ref{eq6}) and the equilibrium
concentration (\ref{eq7}) of the electron-hole liquid in the CQW do not
depend on the parameter $\nu .$ This is in spite of the fact that formally
this parameter was supposed to be large. This circumstance allows us to
expect that the liquid equilibrium electron-hole state in the CQW may arise
even for $\nu \sim 1.$ However, for $\nu \gg 1,$ the kinetic energy, which
is of the order of $\varepsilon _{F}\sim \frac{1}{\nu },$ plays an
insignificant role in the formation of the equilibrium state. The
equilibrium within the electron-hole system is reached due to balance
between the positive direct Coulomb interaction (\ref{eq2.2111}) and the
negative \textit{in-layer} correlation energy (\ref{eq4.31}).

Let us pay attention to the interesting feature of the origin for the
correlation energy associated with the RPA diagram. The correlation energy
contribution results from the momentum transfer $p\sim n^{1/3}\gg p_{F}.$
This conclusion strongly differs from that of Gell-Mann and Brueckner who
considered the many-body correlations for a $3D$ electron gas. Within their
approach, the correlation energy originates from the momentum transfer $p\ll
p_{F}.$ Let us also point out additional arguments which approve the RPA
approximation. While the Gell-Mann and Brueckner approach is justified for $%
n\gg 1,$ our approach holds for an arbitrary concentration $n$ provided that
$\nu $ is large.

Let us investigate how the results obtained can be associated with the
experiments. Below, we follow the scenario proposed by Keldysh for the
electron-hole liquid formation in semiconductors\cite{keldysh}. Consider a
non-equilibrium inhomogeneous electron-hole system in the CQW, composed of
two coexisting phase: a rare exciton gas phase and a dense electron-hole
liquid drops. Let $R$ be the characteristic radius of the drop and $\tau $
be the carrier (electron or hole) lifetime within this drop. Then, the loss
of the carriers per unit time for the drop is $-\pi R^{2}n_{eq}/\tau .$ On
the other hand, there exists an exciton flow $j$ in the gas phase, induced
by an external source. This flow gains the number of carriers in the drop
per unit time $+2\pi Rj.$ In the dynamic equilibrium, the loss and the gain
balance each other, resulting in the dependence%
\begin{equation*}
R\approx \frac{2j\tau }{n_{eq}}.
\end{equation*}%
The concept of the electron-hole drop in the CQW fails if $R<a_{B}=1.$%
Therefore, there exists a threshold value for the flow%
\begin{equation}
j>j_{c}=\frac{n_{eq}a_{B}}{2\tau },  \label{eq9}
\end{equation}%
which provides the formation of the drop. Thus, if the flow $j$ is weak, the
drop has no time to form.

Such a scenario may account for the appearance of luminescent rings in the
CQW in the series of experiments \cite{Nature-2002-Butov, Nature-2002-Snoke,
Butov2003, Snoke2003, Butov2004, Larionov2002, Dremin2004, Timofeev2005}.
These rings, sometimes, are associated with the exciton Bose condensate.
However, in our opinion, they should be connected with the electron-hole
liquid drop described above. First of all for these experiments, the
criterion of the exciton existence $R_{ex}^{2}n\ll 1$ seems to be violated.
However, suppose that the core of the ring under the experiments is
illuminated by the external optical pump resulting in the creation of the
exciton gas flow $j$ within the core. Then, the ring corresponds to the
electron-hole liquid phase described above. It follows from Eq. (\ref{eq9})
that the electron-hole liquid ring can dwell in the stationary state if the
flow $j$ exceeds a certain critical value. In the experiments, the
luminescent ring appears if the optical pumping exceeds a certain power.
Thus, the results obtained may be the basis for the explanation of these
experiments.

Let us mention that the correlation energy calculated above does not involve
the effect of the exciton-like electron-hole correlation associated with the
ladder diagrams in the electron-hole scattering channel. The involvement of
this effect would result in the appearance of the nonvanishing anomalous
exciton-like averages responsible for the creation of the dielectric gap on
the Fermi surface\cite{KO}. However, the contribution of these diagrams into
correlation energy is small in the parameters $1/\nu .$This makes
insignificant their influence on the thermodynamics of the system, the main
subject of the paper.

We are grateful to Yuri Lozovik for valuable discussions.


\end{document}